\documentclass[aps,pra,showpacs,showkeys,onecolumn]{revtex4}
\usepackage{graphicx}
\usepackage{latexsym}
\usepackage{makeidx}
\usepackage{amsmath}
\usepackage{amssymb}
\usepackage{graphicx}

\input{tcilatex}

\begin{document}

\title{Photon creation from vacuum and interactions engineering in \\
nonstationary circuit QED}



\author{A V Dodonov}

\affiliation{Departamento de F\'{\i}sica, Universidade Federal de
S\~{a}o Carlos, 13565-905, S\~ao Carlos,
SP, Brazil}

%\eads{\mailto{adodonov@df.ufscar.br} }

\begin{abstract}
We study theoretically the nonstationary circuit QED system in which the
artificial atom transition frequency, or the atom-cavity coupling, have a
small periodic time modulation, prescribed externally. The system formed by
the atom coupled to a single cavity mode is described by the Rabi
Hamiltonian. We show that, in the dispersive regime, when the modulation
periodicity is tuned to the `resonances', the system dynamics presents the
dynamical Casimir effect, resonant Jaynes-Cummings or resonant
Anti-Jaynes-Cummings behaviors, and it can be described by the corresponding
effective Hamiltonians. In the resonant atom-cavity regime and under the
resonant modulation, the dynamics is similar to the one occurring for a
stationary two-level atom in a vibrating cavity, and an entangled state with
two photons can be created from vacuum. Moreover, we consider the situation
in which the atom-cavity coupling, the atomic frequency, or both have a
small nonperiodic time modulation, and show that photons can be created
from vacuum in the dispersive regime. Therefore, an analog of the dynamical
Casimir effect can be simulated in circuit QED, and several photons, as well
as entangled states, can be generated from vacuum due to the anti-rotating
term in the Rabi Hamiltonian.
\end{abstract}

\maketitle

\section{Introduction}

Over the last few decades nonstationary processes in cavities have received
considerable attention. One such process is the nonstationary or dynamical
Casimir Effect (DCE) -- in particular, the creation of photons from vacuum,
or another initial field state, in cavity whose geometry \cite%
{PRA50-1027,PRA50-1027a,PRA50-1027b,Law,Law1} or material properties \cite%
{Law,Law1,jj,jj1,jj2,jj3} have a periodic time dependence, with the
modulation frequency close to twice the unperturbed field eigenfrequency.
Nowadays, DCE in cavities is a well studied problem, with a variety of
theoretical predictions concerning the number and the statistics of created
photons, as well as the influence of detuning, dissipation, boundary
conditions, geometry and nonperiodicity of the modulation (see \cite%
{JPB-S,book,JRLR26-445} for an extensive list of references). To date, DCE
has not been observed in laboratory, however several concrete proposals have
appeared over the last years \cite%
{JPA39-6271,JPA39-6271a,onofrio,JPA39-6271b,d2,mme}, with some of them being
currently implemented experimentally \cite{EPL70-754,EPL70-754a,braggio}.

The interest in nonstationary processes in cavities reappeared over the last
5 years due to the progress in the field of Cavity Quantum Electrodynamics
(cavity QED \cite{S298-1372,S298-1372a}) in the condensed matter systems,
e.g. semiconductor quantum dots \cite%
{N432-197,N432-200,faraon,faraon1,ima,ima1}, polar molecules \cite%
{NP2-636,NP2-636a} and superconducting circuits \cite%
{N431-159,N431-162,PRL96-127006} coupled to high-$Q$ resonators, the latest
architecture known as \textit{circuit} QED \cite{N431-162,revvv}. In cavity
QED, the effective 2-level atom is coupled to the field inside the resonator
via the dipole interaction, allowing for observation of the light-matter
interaction at the level of single photons and single atoms. The new
ingredient in the solid state cavity QED, in particular circuit QED, is the
possibility of engineering and manipulating the properties of the artificial
2-level atom and the resonator \cite{N431-162,xx5,xx6}, as well as the
interaction strength between them \cite{PRA69-062320}, either during the
fabrication, or \textit{in situ}.

Recently, the strong resonant and the strong dispersive coupling limits
between the artificial atom and a single cavity mode were observed
experimentally in circuit QED \cite{N431-162,N445-515} and other solid state
architectures \cite{N432-197,N432-200,faraon,faraon1,ima1}. Moreover, the
single photon source \cite{N449-7160}, single artificial-atom maser \cite%
{xx4}, multiphoton Fock states \cite{medv} and interaction between two
artificial atoms (qubits) \cite{xx6,N449-443} were implemented
experimentally, among many other important achievements (see \cite{revvv}
for more references). Besides, the circuit QED architecture benefits from
robust readout schemes of the atomic and the resonator states \cite%
{PRA69-062320,medv,r1,r1a,r1b,r2,PRA75-032329}, relatively low dissipative
losses \cite{N445-515}, state preparation techniques \cite{gp} and real time
manipulation of the atomic transition frequency via electric and magnetic
fields \cite{N431-162,xx6,medv} or nonresonant microwave fields \cite%
{N449-443,PRA75-032329}.

Harnessing the tunability of the atomic transition frequency, the authors in
\cite{lz,lz1} proposed an experimental implementation of the Landau-Zener
sweeps in circuit QED, when the atomic frequency increases linearly in time.
This allows for generation of single photons, as well as cavity-atom
entangled states. In a similar direction, the generation of the quantum
vacuum radiation by modulating periodically the vacuum Rabi frequency of an
intersubband transition in a doped quantum well system embedded in a planar
microcavity was considered in \cite{d2,mme,d1,PRB72-115303}, and emission of
photon pairs was predicted for `resonant' modulation frequencies. A
preliminary theoretical study of the feasibility of realizing the dynamical
Casimir effect with a quantum flux qubit in superconducting quantum
nanocircuits, as well as the detection of the generated photons, was
reported in \cite{sss}. On the other hand, the freedom of controlling in
real-time the atomic frequency is currently being explored to
couple/decouple one or several qubits to/from the cavity mode in order to
implement quantum logic operations \cite{xx6,N449-443,PRA75-032329}.

In this paper, following the original proposal \cite{I-lukin}, we study
nonstationary processes in the solid state cavity QED, where a single cavity
mode is coupled to a single artificial atom whose transition frequency, or
atom-cavity coupling, have a small periodic time modulation, prescribed
externally. Such a control over the transition frequency, with compatible
modulation frequency, can be achieved in circuit QED with present or
near-future technology \cite{xx6,medv,PRA75-032329}. We show that in the
dispersive atom-cavity regime and under the resonance conditions one obtains
completely different effective regimes of the atom-field interaction, which
can be approximately described by the resonant Anti-Jaynes-Cummings (AJC),
resonant Jaynes-Cummings (JC) or the dynamical Casimir effect (DCE)
Hamiltonians. Moreover, in the resonant atom-cavity regime, the system
dynamics resembles the behavior of the DCE in the vibrating cavity
containing a resonant two-level atom, and entangled states with at most two
photons can be generated from vacuum.

We also consider the case, in which the atomic frequency, the atom-cavity
coupling parameter, or both have a small nonperiodic time modulation,
prescribed externally. Namely, we suppose that the modulation is given by
the sum of two harmonic functions with different amplitudes and frequencies.
We deduce an effective Hamiltonian in the dispersive regime, and show that
photon generation from vacuum is possible for fine tuned modulation
frequencies.

Thus, we demonstrate the possibility of simulating the DCE in circuit QED
using a single nonstationary atom, instead of a macroscopic dielectric
medium as in Refs. \cite{jj,jj1,jj2,jj3}. As applications, it could be
possible to create excitations, either photonic or atomic, from the initial
vacuum state $|g,0\rangle $, generate nonclassical states of light and
realize transitions between the states $\left\{ |g,n\rangle ,|e,n\pm
1\rangle \right\} $ in the dispersive regime, where $|g\rangle $ and $%
|e\rangle $ are the atomic ground and excited states, respectively, and $%
|n\rangle $ is the Fock state of the cavity field. A related problem was
recently studied in \cite{today}, where it was suggested that lasing
behavior and the creation of a highly non-thermal population of the
oscillator, as well as the cooling, could be implemented using an analogous
scheme in the near future.

\section{Nonstationary circuit QED with periodic modulation}

We assume that the atomic transition frequency $\Omega \left( t\right) $ is
given by%
\begin{equation}
\Omega \left( t\right) =\Omega _{0}+\varepsilon f_{t}\,.
\end{equation}
Here $\Omega _{0}$ denotes the bare atomic frequency, $\varepsilon $ is a
small modulation amplitude, $\varepsilon \ll \Omega _{0}$, and $f_{t}$ is an
arbitrary periodic function of time, prescribed externally%
\begin{equation}
f_{t}=\sum_{k=0}^{\infty }\left[ s_{k}\sin k\eta t+c_{k}\cos k\eta t\right] ,
\label{ft}
\end{equation}%
where $\eta $ is the modulation frequency and $\left\{ s_{k},c_{k}\right\} $
form a set of coefficients describing the time modulation. The cavity
frequency $\omega $ and the atom-cavity coupling parameter $g_{0}$ are
constant, so at the charge degeneracy point \cite{PRA69-062320} the system
is described by the Rabi Hamiltonian (RH)
\begin{equation}
H=H_{0}\left( t\right) +g_{0}(a+a^{\dagger })\left( \sigma _{+}+\sigma
_{-}\right) ,  \label{Rabi}
\end{equation}%
with $a$ ($a^{\dagger }$) being the cavity annihilation (creation) operator,
$\sigma _{+}=|e\rangle \langle g|$ and $\sigma _{-}=|g\rangle \langle e|$.
The free Hamiltonian is%
\begin{equation}
H_{0}\left( t\right) =\omega \hat{n}+\frac{\Omega \left( t\right) }{2}\sigma
_{z},
\end{equation}
where $\hat{n}=a^{\dagger }a$ is the photon number operator, $\sigma
_{z}=|e\rangle \langle e|-|g\rangle \langle g|$ and we assume $\hbar =1$. In
the stationary case, $\varepsilon =0$, one can perform the Rotating Wave
Approximation (RWA) and obtain the standard Jaynes-Cummings (JC) Hamiltonian
\cite{Schleich}, which has been verified in several experiments over the
last few years \cite{N431-162,N445-515,climbing,rempe,ss}. However, in the
nonstationary case, as well as under strong dephasing noise \cite{TW}, the
anti-rotating term $(a\sigma _{-}+a^{\dagger }\sigma _{+})$ cannot always be
eliminated. Moreover, it is responsible for producing an analog of the DCE
and creating photonic and atomic excitations from vacuum under modulation
`resonance' conditions, as shown below.

In the interaction picture with respect to $H_{0}\left( t\right) $ the
interaction Hamiltonian reads
\begin{equation}
H_{I}=g_{0}\left( e^{i\Xi _{-}}a\sigma _{+}+e^{i\Xi _{+}}a^{\dagger }\sigma
_{+}+h.c.\right) ,  \label{a1}
\end{equation}%
where h.c. stands for the Hermitian conjugate and
\begin{equation}
\Xi _{\pm }\equiv \int_{0}^{t}d\tau \left[ \Omega \left( \tau \right) \pm
\omega \right] .
\end{equation}
All the information about the influence of the external modulation on the
system dynamics is contained in the time-dependent coefficients $\exp (i\Xi
_{\pm })$, which can be significantly simplified by tuning the modulation
frequency $\eta $ to the `resonances'. We have explicitly
\begin{equation}
g_{0}e^{i\Xi _{\pm }}=ge^{i\Delta _{\pm }t}\sum_{l=0}^{\infty }\frac{1}{l!}%
\left[ \frac{\varepsilon }{\eta }\sum_{k=1}^{\infty }\left( \Lambda
_{k}e^{-ik\eta t}-\Lambda _{k}^{\ast }e^{ik\eta t}\right) \right] ^{l},
\label{eee}
\end{equation}%
where we defined the complex coupling constant
\begin{equation}
g\equiv g_{0}\exp \left[ i\frac{\varepsilon }{\eta }\sum_{k=1}^{\infty
}k^{-1}s_{k}\right]
\end{equation}
and parameters%
\begin{equation}
\Lambda _{k}\equiv -\frac{c_{k}+is_{k}}{2k}
\end{equation}
\begin{equation}
\Delta _{\pm }\equiv (\Omega _{0}+\varepsilon c_{0})\pm \omega .
\label{Delty}
\end{equation}

\section{Anti-Jaynes-Cummings (AJC) resonance}

The `Anti-Jaynes-Cummings' (AJC) resonance occurs for%
\begin{equation}
\eta =\eta _{AJC}\equiv \Delta _{+}-\xi ,
\end{equation}
where $\left\vert \xi \right\vert \ll \Delta _{+}$ is a small `resonance
shift'\ to be adjusted afterwards. Assuming a reasonable experimental
condition $\varepsilon /\eta \ll 1$, we expand $g_{0}\exp (i\Xi _{\pm })$ in
(\ref{a1}) to the first order in $\varepsilon /\eta $ and make RWA,
obtaining
\begin{equation}
H_{I}\simeq g\left( \theta e^{i\xi t}a^{\dagger }\sigma _{+}+e^{i\Delta
_{-}t}a\sigma _{+}\right) +h.c.,  \label{resul}
\end{equation}%
where the modulation induced dimensionless coupling is
\begin{equation}
\theta \equiv \Lambda _{1}\frac{\varepsilon }{\eta }\,.  \label{theta}
\end{equation}

\subsection{Dispersive regime}

In the dispersive regime, $g_{0}\sqrt{\left\langle \hat{n}\right\rangle }%
/\left\vert \Delta _{-}\right\vert \ll 1$, where $\left\langle \hat{n}%
\right\rangle $ is the mean photon number, the Hamiltonian (\ref{resul}) can
be approximated by \cite{Schleich}%
\begin{equation}
H_{I}^{\left( 1\right) }\simeq \left( g\theta e^{i\xi t}a^{\dagger }\sigma
_{+}+h.c.\right) +\delta \left( \hat{n}+1/2\right) \sigma _{z}~,
\label{disc}
\end{equation}%
where
\begin{equation}
\delta =\delta _{0}+\mathcal{O}\left( g_{0}^{2}/\Delta _{+}\right)
\label{disc1}
\end{equation}%
and%
\begin{equation}
\delta _{0}=\frac{g_{0}^{2}}{\Delta _{-}}
\end{equation}%
is the standard dispersive shift \cite{Schleich}. After the unitary
time-dependent transformation defined by the operator
\begin{equation}
U=\exp \left[ i\xi t\sigma _{z}/2\right]
\end{equation}%
we get the effective AJC Hamiltonian
\begin{equation}
H_{AJC}=U^{-1}H_{I}^{\left( 1\right) }U-iU^{-1}\frac{\partial U}{\partial t}%
\simeq \frac{\xi +\delta \left( 1+2\hat{n}\right) }{2}\sigma _{z}+\left(
g\theta a^{\dagger }\sigma _{+}+h.c.\right) .  \label{AJC}
\end{equation}

For the field initially in the Fock state, by adjusting the resonance shift $%
\xi $ in order to make $\left\vert \xi +\delta \left( 1+2\hat{n}\right)
\right\vert $ small compared to $\left\vert g\theta \right\vert $, one
obtains approximately the resonant AJC Hamiltonian
\begin{equation}
H_{AJC}^{(r)}\simeq g\theta a^{\dagger }\sigma _{+}+g^{\ast }\theta ^{\ast
}a\sigma _{-}.  \label{AJC-r}
\end{equation}%
From the physical point of view, the external modulation supplies the energy
$\omega +\Omega _{0}$ necessary to create one photon and one atomic
excitation simultaneously. Thus, one can create the superposition of states $%
|e,1\rangle $ and $|g,0\rangle $ starting from the initial vacuum state $%
|g,0\rangle $ and fine tuning the resonance shift.

\begin{figure}[tbh]
\begin{center}
\includegraphics[width=.6\textwidth]{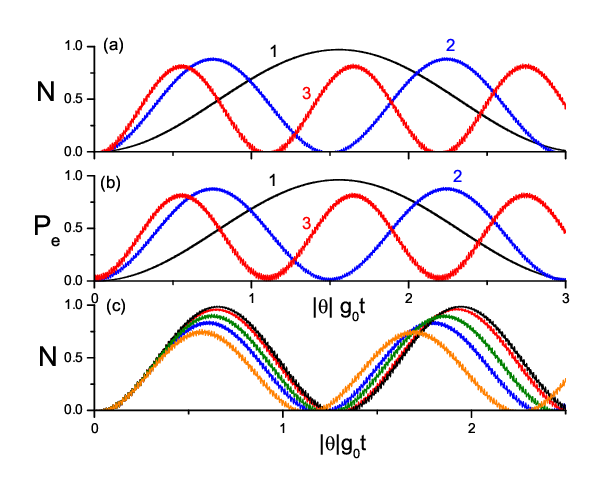}
\end{center}
\caption{The results of the numerical integration of the Schr\"{o}dinger
equation with the Rabi Hamiltonian (\protect\ref{Rabi}) for the modulation
frequency $\protect\eta =\Delta _{+}+2\protect\delta _{0}\left(
1+n_{0}\right) $. We see an AJC behavior, as given by the effective
Hamiltonian (\protect\ref{AJC}). \textbf{a)} The number of created photons $%
N $ and \textbf{b)} the atom excitation probability $P_{e}$ versus the
dimensionless time for the initial Fock states $|g,n_{0}\rangle $, with $%
n_{0}=0$ (line 1), $n_{0}=3$ (line 2) and $n_{0}=6$ (line 3). \textbf{c)}
The dependence of $N$ on the small change $x$ in the resonance shift $%
\protect\xi =-\protect\delta \left( 1+n_{0}+x\right) $ for the initial
number state $|g,n_{0}=5\rangle $. For $\left\vert \protect\theta %
\right\vert g_{0}t$ from 2 to 2.3 the curves correspond to $x=4,0,3,1,2$,
from below. By adjusting $x$ one can optimize the photon generation process.}
\label{Fig1}
\end{figure}

In figure \ref{Fig1} we show the exact dynamics for the AJC resonance,
obtained through the numerical integration of the Schr\"{o}dinger equation
with the initial Hamiltonian (\ref{Rabi}), setting the experimental circuit
QED parameters $\Omega _{0}/\omega =1.4$, $g_{0}/\omega =2\cdot 10^{-2}$ and
assuming $\varepsilon /\omega =0.2$. In all the simulations throughout this
paper we consider the harmonic modulation, $f_{t}=\sin \eta t$. In the
figure \ref{Fig1}a we plot the number of created photons
\begin{equation}
N\equiv \left\langle \hat{n}\right\rangle -n_{0},
\end{equation}
for the initial Fock state $|g,n_{0}\rangle $ as function of the
dimensionless time $\left\vert \theta \right\vert g_{0}t\,$, assuming the
resonance shift $\xi =-2\delta _{0}\left( 1+n_{0}\right) $. In figure \ref%
{Fig1}b we plot the atom excitation probability $P_{e}$. From figures \ref%
{Fig1}a and \ref{Fig1}b one can observe the simultaneous generation of one
photon and one atomic excitation, in agreement with the effective AJC
Hamiltonian (\ref{AJC}). However, the maximal mean number of created photons
is slightly less than $1$ and depends on the initial photon number $n_{0}$,
meaning that the resonance shift was not adjusted precisely. In figure \ref%
{Fig1}c we plot $N$ for the initial number state $|g,n_{0}=5\rangle $ for
different resonance shifts $\xi =-2\delta _{0}\left( 1+n_{0}+x\right) $, $%
x=0,1,2,3,4$, demonstrating that it is possible to find an optimum $\xi $
which reproduces the resonant AJC Hamiltonian and allows for the creation of
one photon and the atomic excitation simultaneously.

\subsection{Resonant regime}

In the resonant regime, $\left\vert \Delta _{-}\right\vert /g_{0}\ll 1$, the
effective Hamiltonian method is not applicable. So we apply the method of
slowly varying amplitudes on the Hamiltonian (\ref{resul}), repeating the
procedure \cite{pp} employed originally for studying the photon generation
from vacuum due to the DCE in a vibrating cavity containing a resonant
(stationary) two-level atom \cite{Lozovik,Lozovik1,Lozovik2}. We find that
for the initial state $|g,0\rangle $ the photon generation occurs for two
possible resonance shifts%
\begin{equation}
\xi _{\pm }=\Delta _{-}/2\pm \sqrt{2}g_{0}
\end{equation}
and one gets the following non-zero probabilities $P_{an}$, with $a=\left(
e,g\right) $ denoting the atomic state and $n$ the photon number%
\begin{eqnarray}
P_{g0} &\approx &\cos ^{2}\left( \chi t\right)  \nonumber \\
P_{e1} &\approx &\sin ^{2}\left( y+q\right) \sin ^{2}\left( \chi t\right)
\label{PPP} \\
P_{g2} &\approx &\cos ^{2}\left( y-q\right) \sin ^{2}\left( \chi t\right) .
\nonumber
\end{eqnarray}%
Here%
\begin{equation}
\chi \approx \left\vert g\theta \right\vert \,\sin \left( y+q\right) ,\quad
\tan y\approx \left( \frac{ 2\sqrt{2}g_{0}+\Delta _{-} }{ 2\sqrt{2}%
g_{0}-\Delta _{-} }\right) ^{1/2}
\end{equation}
and $q=0$ ($\pi /2$) for $\xi _{-}$ ($\xi _{+}$).
\begin{figure}[htb]
\begin{center}
\includegraphics[width=.5\textwidth]{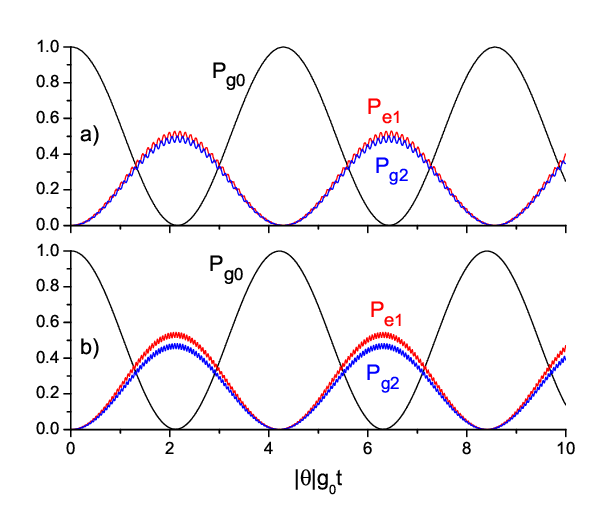} {}
\end{center}
\caption{The time dependence of the probabilities $P_{g0}$, $P_{e1}$ and $%
P_{g2}$, obtained by the numerical integration of the Schr\"{o}dinger
equation with the Rabi Hamiltonian (\protect\ref{Rabi}) in the resonant
atom-cavity regime for the AJC resonance with properly adjusted frequency
shift. \textbf{a) } $\Delta _{-}=0$, $\protect\xi =\protect\xi _{-}$ and $%
\protect\varepsilon /\protect\omega _{r}=2\cdot 10^{-1}$. \textbf{b)} $%
\Delta _{-}=10^{-1}g_{0}$, $\protect\xi =\protect\xi _{-}$ and $\protect%
\varepsilon /\protect\omega _{r}=10^{-1}$. Not more than $2$ photons can be
generated from vacuum, and the atom excitation probability is limited by the
value $\sim 1/2$, in agreement with equation (\protect\ref{PPP}). }
\label{Fig2}
\end{figure}

We illustrate this behavior in figure \ref{Fig2}, where we show $P_{g0}$, $%
P_{e1}$ and $P_{g2}$ in the resonant atom-cavity regime under the AJC
resonance. In figure \ref{Fig2}a we consider $\Delta _{-}=0$ and $\xi =\xi
_{-}$, using the parameters $g_{0}/\omega =4\cdot 10^{-2}$ and $\varepsilon
/\omega =2\cdot 10^{-1}$. In figure \ref{Fig2}b we set $\Delta
_{-}=10^{-1}g_{0}$, $\xi =\xi _{-}$, $g_{0}/\omega =4\cdot 10^{-2}$ and $%
\varepsilon /\omega =10^{-1}$. The results are in excellent agreement with
the theoretical predictions, equation (\ref{PPP}), demonstrating that a
superposition of states $|g,0\rangle $, $|e,1\rangle $ and $|g,2\rangle $ is
created from the initial vacuum state $|g,0\rangle $. This dynamics is
similar to the one studied in the context of DCE, where a resonant
(stationary) two-level atom is fixed inside a cavity whose boundary is
oscillating with the frequency close to $2\omega $ \cite{pp}. In both cases
not more than two photons can be created from the vacuum state $|g,0\rangle $
and the probability of exciting the atom is limited by the value $1/2$. This
similarity is not surprising, because for the AJC resonance the modulation
frequency is $\eta \approx \Delta _{+}\approx 2\omega $, and the artificial
atom plays the role of the stationary two-level atom and the cavity
modulating mechanism at the same time.

\section{Jaynes-Cummings (JC) resonance}

In the dispersive regime the `Jaynes-Cummings' (JC) resonance occurs for%
\begin{equation}
\eta =\eta _{JC}\equiv \left\vert \Delta _{-}\right\vert -\xi .
\end{equation}
For $\Delta _{-}>0$, repeating the steps that led us to (\ref{AJC}), we get
the effective JC Hamiltonian%
\begin{equation}
H_{JC}\simeq \frac{\xi +\delta \left( 1+2\hat{n}\right) }{2}\sigma
_{z}+\left( g\theta \,\,a\sigma _{+}+h.c.\right) .  \label{JC}
\end{equation}%
If $\Delta _{-}<0$, we obtain (\ref{JC}) upon the replacements $\theta
\rightarrow -\theta ^{\ast }$ and $\xi \rightarrow -\xi $. Thus, by
employing the JC resonance and adjusting the resonance shift $\xi $, one can
couple the subspaces $\left\{ |g,n\rangle ,|e,n-1\rangle \right\} $ when the
atom and the field are far off resonant, since the external modulation
supplies the energy difference $|\Omega_0-\omega|$ necessary to couple the
atom and the cavity field.
\begin{figure}[tbh]
\begin{center}
\includegraphics[width=.6\textwidth]{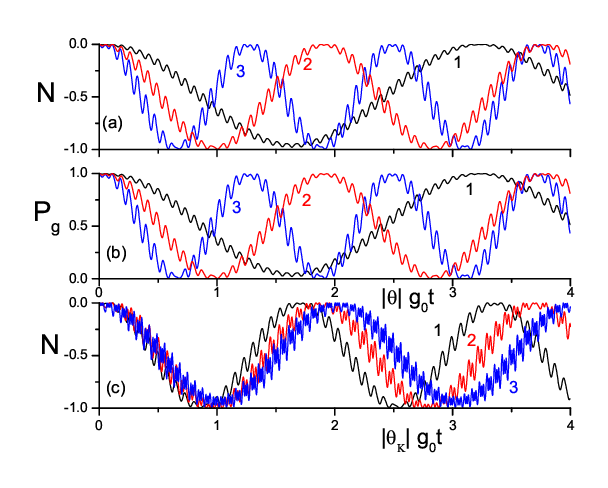} {}
\end{center}
\caption{The results of the numerical integration of the Schr\"{o}dinger
equation with the Rabi Hamiltonian (\protect\ref{Rabi}) for the JC
resonance, $\protect\eta =\Delta _{-}+2\protect\delta _{0}\left(
1+n_{0}\right) $, considering initial number states $|g,n_{0}\rangle $ with $%
n_{0}=1$ (line 1), $n_{0}=3$ (line 2) and $n_{0}=7$ (line 3). \textbf{a)}
The number of created photons $N$ and \textbf{b)} the probability of finding
the atom in the ground state $P_{g}$ as functions of the dimensionless time
demonstrate a transfer between the states $|g,n_{0}\rangle \ $and $%
|e,n_{0}-1\rangle $ in the dispersive regime. \textbf{c)} $N$ for the
initial number state $|g,4\rangle $ and three lower JC resonances ($\protect%
\eta =\protect\eta _{-}^{\left( K\right) }$, $K=1,2,3$ as shown, see section
\protect\ref{zzz}) for the resonance shift $\protect\xi =-10\protect\delta %
_{0}$. }
\label{Fig3}
\end{figure}

The JC behavior is illustrated in figure \ref{Fig3} for the initial Fock
state $|g,n_{0}\rangle $. In figures \ref{Fig3}a and \ref{Fig3}b we plot $N$
and $P_{g}=1-P_{e}$, respectively, using the same parameters as in figure %
\ref{Fig1} and setting $\eta =\Delta _{-}+2\delta _{0}\left( 1+n_{0}\right) $%
. As expected from the effective JC Hamiltonian (\ref{JC}) one has
transitions between the states $\left\{ |g,n\rangle ,|e,n-1\rangle \right\} $
for large detuning, $\left\vert \Delta _{-}\right\vert \gg g$. Thus, in the
dispersive regime one could engineer entangled states with several photons
from the initial vacuum state $|g,0\rangle $ by alternating between the AJC
and JC resonances and controlling the time interval of each resonance.

\section{Dynamical Casimir effect (DCE) resonance}

In the dispersive regime, the `dynamical Casimir effect' (DCE) resonance
occurs for%
\begin{equation}
\eta =\eta _{DCE}\equiv \Delta _{+}-\Delta _{-}-2\xi =2\omega -2\xi .
\end{equation}%
Performing RWA in the interaction Hamiltonian (\ref{a1}) we obtain to the
first order in $\varepsilon /\eta $%
\begin{equation}
H_{I}\simeq ge^{i\Delta _{-}t}a\sigma _{+}+g\theta e^{i\left( \Delta
_{-}+2\xi \right) t}a^{\dagger }\sigma _{+}+h.c.
\end{equation}%
Applying the unitary time-dependent transformation defined by the operator
\begin{equation}
U_{1}=\exp \left\{ i\left[ \xi \hat{n}+\left( \Delta _{-}+\xi \right) \sigma
_{z}/2\right] t\right\} ,
\end{equation}%
we obtain the time-independent Hamiltonian consisting of the JC Hamiltonian
plus the anti-rotating term multiplied by the adjustable coupling $g\theta $%
\begin{equation}
H_{I}^{\left( 1\right) }\simeq \xi \hat{n}+\frac{\Delta _{-}+\xi }{2}\sigma
_{z}+\left( ga\sigma _{+}+g\theta a^{\dagger }\sigma _{+}+h.c.\right) ~.
\end{equation}%
Since $\left\vert g/\Delta _{-}\right\vert \ll 1$, we can obtain an
effective Hamiltonian by applying a sequence of small unitary
transformations on $H_{I}^{\left( 1\right) }$ \cite{Klimov} and performing
the Hausdorff expansion in some small parameter at each step. Here our
expansion parameter is $\epsilon =\left\vert g/\Delta _{-}\right\vert $ and,
assuming that $\theta \sim \mathcal{O}\left( \epsilon \right) $, we consider
terms up to the second order in $\epsilon $. Applying first the `rotating'
unitary transformation defined by the time-independent operator
\begin{equation}
U_{r}=\exp \left[ \left( ga\sigma _{+}-h.c.\right) /\Delta _{-}\right]
\end{equation}%
followed by the `antirotating' transformation defined by the operator%
\begin{equation}
U_{a}=\,\exp \left[ \left( g\theta a^{\dagger }\sigma _{+}-h.c.\right)
/\left( \Delta _{-}+2\xi \right) \right] \,,
\end{equation}%
we obtain the effective Hamiltonian
\begin{equation}
H_{eff}=U_{a}U_{r}H_{I}^{\left( 1\right) }U_{r}^{\dagger }U_{a}^{\dagger }
\end{equation}%
of the following form (considering terms up to the second order in $\epsilon
$)%
\begin{equation}
H_{eff}\simeq \left( \xi +\delta \sigma _{z}\right) \hat{n}+\delta \sigma
_{z}\left( \theta ^{\ast }a^{2}+h.c.\right) +\frac{\Delta _{-}+\delta +\xi }{%
2}\sigma _{z}-\frac{2\delta }{\Delta _{-}}\left( ga\hat{n}\sigma
_{+}+h.c.\right) +\mathcal{O}\left( \epsilon ^{3}\right) .
\end{equation}%
Finally, moving to a rotating frame by means of the unitary time-dependent
transformation defined by the operator
\begin{equation}
U_{2}=\exp \left[ -i\left( \Delta _{-}+\delta +\xi \right) \sigma _{z}t/2%
\right] ,
\end{equation}%
we obtain
\begin{equation}
H_{DCE}\simeq \left( \xi +\delta \sigma _{z}\right) \hat{n}+\delta \sigma
_{z}(\theta a^{\dagger 2}+\theta ^{\ast }a^{2})-\frac{2\delta }{\Delta _{-}}%
\left( ge^{i\Delta _{-}t}a\hat{n}\sigma _{+}+h.c.\right) +\mathcal{O}\left(
\epsilon ^{3}\right) ,  \label{PDC}
\end{equation}%
where we neglected $(\delta +\xi )$ compared to $\Delta _{-}$ in the
exponential of the third term.

This is the main result of this section. The first two terms of the
effective Hamiltonian (\ref{PDC}) form the DCE part and the remaining terms
represent corrections, oscillating with a high frequency $\sim \Delta _{-}$.
For small $\left\langle \hat{n}\right\rangle \ll n_{c}\sim \left( \Delta
_{-}/g\right) ^{2}$, the contribution of these corrections is relatively
small and $\sigma _{z}$ becomes approximately a constant. Assuming that
initially the atom is in the ground state, then for $\left\langle \hat{n}%
\right\rangle \ll n_{c}$ one has $\sigma _{z}\approx -1$ and the right-hand
side of equation (\ref{PDC}) transforms into the DCE Hamiltonian \cite%
{Law,Law1,book}%
\begin{equation}
H_{DCE}\simeq \left( \xi -\delta \right) \hat{n}-\delta (\theta a^{\dagger
2}+\theta ^{\ast }a^{2}).
\end{equation}
If the atom is initially in the excited state, a similar effective
Hamiltonian is obtained through the substitution $\delta \rightarrow -\delta
$. Therefore, by adjusting the frequency shift to $\xi =\pm \delta $,
depending on the initial atomic state \footnote{%
If initially the atom is in the superpositions of states $|g\rangle $ and $%
|e\rangle $, we can choose any of the signs to obtain photon generation.
However, the photon generation is optimized if initially the atom is exactly
in $|g\rangle $ or $|e\rangle $.}, we have photon pairs creation from
vacuum, as well as field amplification, due to an analog of the DCE.

In this case the DCE is simulated by the atomic transition frequency
modulation through the strong atom-cavity coupling. However, the photon
generation process is not steady because after several photons have been
created the third and higher order terms become important and the photon
generation is interrupted. Indeed, the third term describes the non-resonant
atomic excitation by means of the absorption of photons from the cavity.
Nevertheless, the Hamiltonian (\ref{PDC}) shows that it is possible to
simulate DCE and generate several photons from vacuum using a single
artificial atom.

\subsection{Qualitative explanation and numerical results}

This phenomenon can be qualitatively understood as follows. In the
dispersive regime the atom acts as an effective nonlinear capacitance \cite%
{PRA69-062320}, pulling the cavity frequency to approximately
\begin{equation}
\omega \left( t\right) \approx \omega +\frac{\sigma _{z}g_{0}^{2}}{\Delta
_{-}+\varepsilon f_{t}}\approx \left( \omega +\sigma _{z}\delta \right)
-\sigma _{z}\delta \frac{\varepsilon }{\Delta _{-}}f_{t}\,.
\end{equation}
Consequently, it is expected that the periodic modulation of $f_{t}$ with
the modulation frequency close to $\eta \approx 2\left( \omega \pm \delta
\right) $ will lead to DCE \cite{book,pp}, for which the photons are
generated as long as the modulation is present. The energy $2\omega $
necessary to create pairs of photons is provided through the atomic
frequency modulation and the resonance shift $\xi $ must be adjusted \footnote{One could adjust the resonance shift adiabatically in time to enhance the resonance.} in
order to get the constructive interference on the cavity field \cite%
{book,JRLR26-445}. However, in our case the atom gets entangled with the
field due to the third term in (\ref{PDC}), by which the atom acquires some
probability of being excited through photon absorption and the photon
generation process cannot continue asymptotically due to the loss of
constructive interference [the first term in (\ref{PDC}) becomes non-zero].
This is different from the usual DCE situation, in which the properties of
the macroscopic linear, lossless and non-dispersive dielectric medium inside
the cavity are modulated \cite{jj,jj1,jj2,jj3} and the field mode does not
get entangled with individual atoms.

\begin{figure}[tbh]
\begin{center}
\includegraphics[width=.7\textwidth]{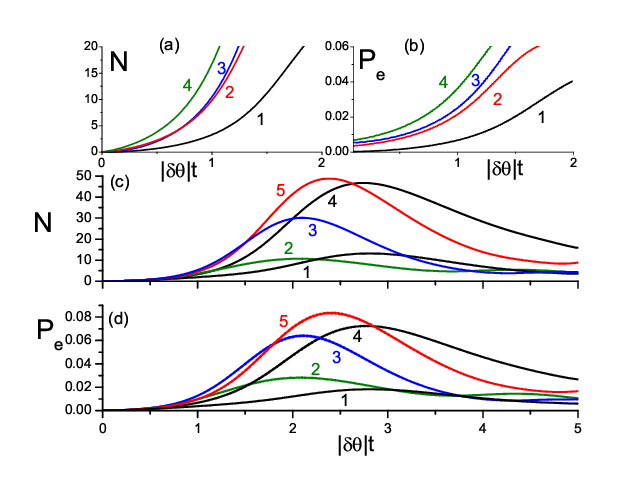} {}
\end{center}
\caption{The initial stage of the time evolution of \textbf{a)} the number
of created photons $N$ and \textbf{b) } the atom excitation probability $%
P_{e}$, obtained by the numerical integration of the Schr\"{o}dinger
equation with the Rabi Hamiltonian (\protect\ref{Rabi}) for $\protect\eta %
=2\left( \protect\omega _{r}-\protect\delta _{0}\right) ,$ for the initial
Fock states $|g,n_{0}\rangle $ with $n_{0}=0$ (line 1), $n_{0}=1$ (line 2)
and coherent states $|g,\protect\alpha \rangle $ with $\left\vert
a\right\vert ^{2}=1/2$ (line 3) and $\left\vert a\right\vert ^{2}=1$ (line
4). There is photon creation due to an analog of the dynamical Casimir
effect, as described by the first two terms in the effective Hamiltonian (%
\protect\ref{PDC}). \textbf{c)} $N$ and \textbf{d)} $P_{e}$ for larger times
and initial state $|g,0\rangle $. Here $\protect\eta =2\left( \protect\omega %
_{r}-x\protect\delta _{0}\right) $ with $x=0.95$ (line 1), $x=1.25$ (line
2), $x=1.15$ (line 3), $x=1$ (line 4) and $x=1.05$ (line 5). The photon
creation is very sensitive to the resonance shift, and the generation
process is interrupted due to the atom-field entanglement, as expected from
the third term in equation (\protect\ref{PDC}).}
\label{Fig4}
\end{figure}

In figure \ref{Fig4} we present the results of numerical integration of the
Schr\"{o}dinger equation with the Rabi Hamiltonian (\ref{Rabi}) for the DCE
resonance with properly adjusted frequency shift, $\eta =2\left( \omega
-\delta _{0}\right) $, using the parameters $\Omega _{0}/\omega
=1.4,g_{0}/\omega =2\cdot 10^{-2},\varepsilon /\omega =4\cdot 10^{-1}$. We
show the number of created photons $N$ (figure \ref{Fig4}a) and the atom
excitation probability $P_{e}$ (figure \ref{Fig4}b, where the fast
oscillations have been averaged out) versus the dimensionless time $%
\left\vert \delta \theta \right\vert t$ for the initial ground atomic state $%
|g\rangle $ and the field number states $|n_{0}\rangle $ ($n_{0}=0,1$) and
coherent states $|g,\alpha \rangle $ ($\left\vert \alpha \right\vert
^{2}=1/2,1$). For initial times, while $\left\langle \hat{n}\right\rangle
\ll n_{c}\sim 400$, there is photon generation and amplification, as
expected. However, the atom gets entangled with the field and $P_{e}$ grows
as time goes on, suffering fast oscillations (not shown for clarity), as
expected from equation (\ref{PDC}).

In figures \ref{Fig4}c and \ref{Fig4}d we show $N$ and $P_{e}$,
respectively, for larger times and different resonance shifts $\xi =x\delta
_{0}$, $x=\left\{ 0.95,1,1.05,1.15,1.25\right\} $, considering the initial
state $|g,0\rangle $. In the first place, we see that the photon generation
is very sensitive to the resonance shift in the vicinity of $\xi =\delta $,
and near the `resonance', $x\approx 1$, there is a substantial photon
creation from vacuum. However, the photon generation process ceases after
several photons have been created and the photon number decreases, although
the photon generation process restarts afterwards\/ (data not shown). $P_{e}$
resembles the behavior of $N$, but remains small for all the times. We also
verified numerically that similar results are obtained for different atomic
initial states, in agreement with equation (\ref{PDC}). Therefore, for $\eta
=2\left( \omega \pm \delta \right) $ one obtains photon generation from
vacuum, as well as photon number amplification, analogously to the dynamical
Casimir effect~\cite{book}.

\section{Discussion of results for periodic modulation}

\label{zzz}

In previous sections we have considered only the first order resonances. In
general, the $K$-th order resonances occurs for an integer $K$ when%
\begin{equation}
\eta =\eta _{i}^{\left( K\right) }\equiv K^{-1}\eta _{i},
\end{equation}%
where $\eta _{i}$ stands for the AJC, JC and DCE resonances. In this case
one recovers the previous results upon substitutions
\begin{equation}
\delta \rightarrow \delta _{K}\quad \mbox{and}\quad \theta \rightarrow
\theta _{K},
\end{equation}%
as can be deduced from equations (\ref{a1}) and (\ref{eee}). For instance,
to obtain $\theta _{K}$, one expands both sums on right-hand side of
equation (\ref{eee}) and make RWA, keeping the terms which oscillate with
the lowest frequency $\xi $. So $\theta _{K}$ contains contributions due to
the non-harmonic shape of the modulation (e.g. $\Lambda _{K}\,\varepsilon
/\eta $), as well as due to the powers of $\left( \varepsilon /\eta \right) $
[e.g. $\left( \Lambda _{1}\,\varepsilon /\eta \right) ^{K}/K!$]. The
effective dispersive shift $\delta _{K}$ is obtained analogously, by first
performing RWA in (\ref{eee}) and keeping terms which oscillate with
frequencies higher than $\xi $, and then calculating the effective
Hamiltonian, as in equations (\ref{disc}) and (\ref{disc1}). So $\delta _{K}$
contains the contribution of many terms besides $\delta _{0}$, among them
the Bloch-Ziegert shift $g^{2}/\Delta _{+}$ \cite{Klimov} and powers of $%
\left( \varepsilon /\eta \right) $. As an example, in figure \ref{Fig3}c we
show $N$ versus $\left\vert \theta _{K}\right\vert g_{0}t$ for the initial
number state $|g,4\rangle $ and the first three JC resonances, considering
the resonance shift $\xi =-10\delta _{0}$. This example illustrates that
higher order resonances are readily accessible. However, for the higher
order resonances the effective dispersive shift $\delta _{K}$ should be
carefully evaluated either numerically or experimentally, otherwise there is
a risk of missing the exact resonance, since $\left\vert \theta
_{K}\right\vert $ becomes smaller and there is less freedom in committing
small errors in the resonance shift $\xi $.

Our results can be directly transposed to the situation where $\Omega \left(
t\right) =\Omega $ is constant and $g_0=g_0\left( t\right) $ has a periodic
time modulation,
\begin{equation}
g_0\left( t\right) =g_{0}+\varepsilon f_{t}.
\end{equation}
In this case, in the interaction picture the interaction Hamiltonian is%
\begin{equation}
H_{I}=\left[(g_{0}+\varepsilon c_{0})-\varepsilon \sum_{k=1}^{\infty
}k\left( \Lambda _{k}^{\ast }e^{ik\eta t}+\Lambda _{k}e^{-ik\eta t}\right) %
\right]\left( e^{i\left( \Omega -\omega \right) t}a\sigma _{+}+e^{i\left(
\Omega +\omega \right) t}a^{\dagger }\sigma _{+}+h.c.\right) .  \label{boris}
\end{equation}%
If one expands $\exp \left( i\Xi _{\pm }\right) $ in equation (\ref{eee}) up
to the first order in $\varepsilon /\eta $, Hamiltonian (\ref{a1}) becomes
equivalent to Hamiltonian (\ref{boris}), so the results obtained above for $%
\Omega \left( t\right) $ also hold for $g_0\left( t\right) $ after making
appropriate substitutions. The main difference between these two cases is
that in the $\Omega \left( t\right) $ case the higher order $K$-th
resonances occur both due to the powers of $\varepsilon /\eta $ and non-zero
coefficients $\Lambda _{K}$, while in the $g_0\left( t\right) $ case the
higher order resonances are due only to non-zero coefficients $\Lambda _{K}$%
. Therefore, for a harmonic modulation, $f_{t}=\sin \eta t$, in the $\Omega
\left( t\right) $ case one has higher order resonances, but in the $%
g_0\left( t\right) $ case there are only the first order resonances.
Finally, if the cavity frequency $\omega $ is modulated periodically, with
constant $\Omega $ and $g_{0}$, the AJC and JC resonances also occur,
besides the well known DCE resonance \cite{pp,Lozovik,Lozovik1,Lozovik2}.

An experimental verification of this scheme seems possible in circuit QED
architecture with superconducting qubits and coplanar waveguide resonators
\cite{PRA69-062320}, where one can adjust the system parameters \emph{in situ%
} via electric and magnetic fields, as demonstrated in several experiments
\cite{N431-162,xx6,N449-443}. Moreover, several schemes to read the cavity
and the atomic states are currently available \cite{PRA69-062320,N445-515,r2}
(or under investigation \cite{PRA75-032329}). The main task would be finding
a way to modulate periodically the atomic transition frequency with a stable
modulation frequency $\eta \sim 10$\thinspace GHz, what is also within an
experimental reach \cite{xx6}. One could also use this scheme to couple $M$
identical qubits (e.g. superconducting 2-level atoms \cite{xx6,N449-443} or
a cloud of polar molecules \cite{NP2-636,NP2-636a}) to the same cavity mode
and modulate the frequency of $M$ atoms simultaneously, since in this case
the effective coupling is increased to $\sqrt{M}g_{0}$.

One important point we did not analyze here is the dissipation and
decoherence of both the artificial atom and the cavity due to the noisy
solid state environment \cite{PRA75-032329,lula}. Recent experiments
achieved experimental values $\left\{ \kappa /\omega <10^{-4},\gamma /\omega
<10^{-3},\gamma _{ph}/\omega <10^{-3}\right\} $ \cite{N445-515}, where $%
\kappa $ is the cavity decay rate, $\gamma $ is the atomic decay rate and $%
\gamma _{ph}$ is the atomic pure dephasing rate. To deal with dissipation in
a qualitative manner, we compare the rates of the photon generation from
vacuum for each resonance to the dissipation rates. We take\ the current
experimental value of the coupling constant $g_{0}/\omega \approx 2\cdot
10^{-2}$ \cite{N445-515} and assume $\varepsilon /\omega \sim \Delta
_{-}/\omega \sim 10^{-1}$ to make the estimative. The photon creation rate
for the first order DCE resonance is roughly $\left\vert \delta \theta
\right\vert /\omega \sim 10^{-4}$ [equation (\ref{PDC})], and for the first
order AJC resonance is roughly $g_{0}\left\vert \theta \right\vert /\omega
\sim 10^{-3}$ [equation (\ref{AJC})]. Both these values are larger than the
dissipation and decoherence rates (or of the same order of magnitude).
Therefore, the photon creation due to modulation of $\Omega \left( t\right) $
or $g_{0}(t)$ seems possible in the future.

\section{A glance at nonperiodic modulation}

\begin{figure}[t]
\begin{center}
\includegraphics[width=.5\textwidth]{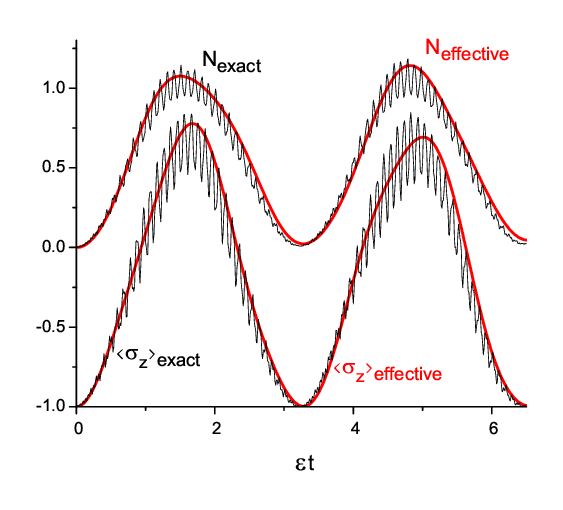} {}
\end{center}
\caption{The mean number of created photons $N$ and atomic population
inversion $\left\langle s_{z}\right\rangle $ obtained numerically for the
exact [equation (\protect\ref{Rabi})] and effective [equation (\protect\ref%
{plus})] Hamiltonians for the resonance shift $\protect\xi =-\protect\delta %
_{0}$. The exact $N$ and $\left\langle \protect\sigma _{z}\right\rangle $
show rapid oscillations, while the effective ones are smooth.}
\label{Fig5}
\end{figure}

Let us now consider that the coupling parameter $g\left( t\right) $ has a
bichromatic external modulation%
\begin{equation}
g\left( t\right) =g_{0}+2\varepsilon \sin [(\Delta _{+}-2\xi
)t]+2\varepsilon _{-}\sin [(\Delta _{-}-2\xi _{-})t],
\end{equation}%
where $g_{0}$ is the bare coupling parameter, $\left\vert \varepsilon
\right\vert ,\left\vert \varepsilon _{-}\right\vert \ll g_{0}$ are small
modulation amplitudes and $\xi ,\xi _{-}$ are the `resonance shifts'. Here $%
\Delta _{\pm }$ is given by equation (\ref{Delty}) with $c_{0}=0$ (i.e., $%
\Delta _{\pm }=\Omega \pm \omega $) and we assume $\Delta _{-}>0.$ The
system Hamiltonian is given by equation (\ref{Rabi}) with time-dependent $%
g\left( t\right) $ and constant $\omega $ and $\Omega $. In the dispersive
regime, we obtain the effective Hamiltonian%
\begin{equation}
H_{eff}\simeq \delta \left( \hat{n}+1/2\right) \sigma _{z}+(i\varepsilon
e^{2i\xi t}a^{\dagger }\sigma _{+}+i\varepsilon _{-}e^{2i\xi _{-}t}a\sigma
_{+}+h.c.).
\end{equation}%
Performing the unitary time-dependent transformation%
\begin{equation}
U_{3}=\exp \left\{ i\left[ \left( \xi +\xi _{-}\right) \sigma _{z}/2+\left(
\xi -\xi _{-}\right) \hat{n}\right] t\right\}
\end{equation}%
we get the time-independent Hamiltonian combining the `rotating' and
`anti-rotating' terms with adjustable couplings%
\begin{eqnarray}
H_{np} &\simeq &U_{3}^{-1}H_{eff}U_{3}-iU_{3}^{-1}\frac{\partial U_{3}}{%
\partial t}  \nonumber \\
&=&\frac{\xi +\xi _{-}+\delta \left( 1+2\hat{n}\right) }{2}\sigma
_{z}+\left( \xi -\xi _{-}\right) \hat{n}+\left( i\varepsilon a^{\dagger
}\sigma _{+}+i\varepsilon _{-}a\sigma _{+}+h.c.\right) .  \label{geral}
\end{eqnarray}

We take equal resonance shifts in order to cancel the second term in (\ref%
{geral}), $\xi =\xi _{-}$, so for $\varepsilon _{-}=0$ ($\varepsilon =0$) we
obtain the effective Anti-Jaynes-Cummings (Jaynes-Cummings) Hamiltonian. A
more interesting regime occurs when both $\varepsilon $ and $\varepsilon
_{-} $ are different from zero. For $\varepsilon =\varepsilon _{-}$ we obtain%
\begin{equation}
H_{+}\simeq \frac{2\xi +\delta \left( 1+2\hat{n}\right) }{2}\sigma
_{z}+i\varepsilon ( a+a^{\dagger }) \left( \sigma _{+}-\sigma _{-}\right) ,
\label{plus}
\end{equation}%
while for $\varepsilon =-\varepsilon _{-}$ we get (\ref{plus}) with the last
term replaced by $i\varepsilon \left( a-a^{\dagger }\right) \left( \sigma
_{+}+\sigma _{-}\right) $. The Hamiltonian (\ref{plus}) cannot be integrated
exactly, so below we solve it numerically and compare the results to the
ones obtained using the initial Rabi Hamiltonian (\ref{Rabi}).

To obtain photon creation from the initial vacuum state $|g,0\rangle $, we
have to make the exponent multiplying $\sigma _{z}$ in (\ref{plus}) as small
as possible by adjusting the resonance shift $\xi $. In figure \ref{Fig5} we
show the mean number of created photons $N$ and the population inversion $%
\left\langle s_{z}\right\rangle $ as function of dimensionless time $%
\varepsilon t$ for $\varepsilon =\varepsilon _{-}$ and $\xi =-\delta _{0}$,
calculated numerically for the initial Hamiltonian (\ref{Rabi}) and the
effective Hamiltonian (\ref{plus}) using the parameters $\Omega /\omega =1.3$%
, $g_{0}/\omega =5\cdot 10^{-2}$ and $\varepsilon /\omega =5\cdot 10^{-3}$.
One can see that the effective Hamiltonian $H_{+}$ describes well the exact
dynamics, and there is a quasi-periodic generation of a few photons,
together with the atomic excitation. Qualitatively, this occurs because for
initial times the Hamiltonian (\ref{plus}) becomes
\begin{equation}
H_{+}\simeq -\sqrt{2}\varepsilon x\sigma _{y},
\end{equation}
and the photons are generated from vacuum quadratically in time. However,
after $\sim 1$ photons have been created, the exponent multiplying $\sigma
_{z}$ in (\ref{plus}) becomes large, so the photon creation process goes out
of resonance and it is interrupted.

\begin{figure}[tbh]
\begin{center}
\includegraphics[width=.5\textwidth]{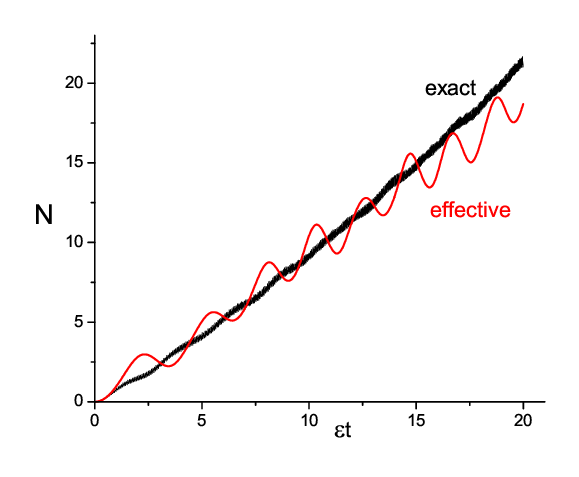} {}
\end{center}
\caption{The mean number of created photons $N$ obtained numerically for the
exact [equation (\protect\ref{Rabi})] and effective [equation (\protect\ref%
{plus})] Hamiltonians for the resonance shift $\protect\xi =\tilde{\protect%
\xi}\left( t\right) $ [equation (\protect\ref{tildz})]. The effective $N$
shows slow oscillations, while the exact one does not.}
\label{Fig6}
\end{figure}

To create a large number of photons from vacuum, we can adopt an `active'
approach, by which the resonance shift $\xi $ is adjusted \emph{adiabatically%
} as function of time, so the effective Hamiltonian (\ref{geral}) is valid.
In this way, we continuously adjust the resonance shift to the number of
photons in the cavity, provided we choose an appropriate functional form of $%
\xi $. In figure \ref{Fig6} we show the mean number of created photons $N$
from the initial state $|g,0\rangle $ for the Rabi Hamiltonian (\ref{Rabi})
and the effective Hamiltonian (\ref{plus}), using the parameters $\Omega
/\omega =1.4$, $g_{0}/\omega =5\cdot 10^{-2}$, $\varepsilon /\omega =5\cdot
10^{-3}$ and setting
\begin{equation}
\xi =\tilde{\xi}\left( t\right) =-\left( \delta _{0}-3g_{0}^{2}/\Delta
_{+}\right) \left( 1/2+\varepsilon t\right) .  \label{tildz}
\end{equation}%
One can see that a significant amount of photons can be generated from
vacuum, provided the resonance shift is adjusted adiabatically, and the
effective Hamiltonian $H_{+}$ describes well the exact dynamics.

Finally, notice that these results also hold if, instead of the coupling
parameter $g$, one modulates the atomic transition frequency $\Omega $, or
both $g$ and $\Omega $ simultaneously, one with frequency $\sim \Delta _{+}$%
, and another with $\sim \Delta _{-}$.

\section{Conclusions}

In conclusion, we studied the nonstationary solid state cavity QED system in
which the atomic transition frequency or the atom-cavity coupling have a
small periodic time modulation, prescribed externally. We showed that in the
dispersive regime and under the resonant modulations, the Rabi Hamiltonian
(which describes the system dynamics) can be significantly simplified,
resulting in three different regimes that can be described by the
Anti-Jaynes-Cummings, Jaynes-Cummings or the dynamical Casimir effect
Hamiltonians. Moreover, in the resonant atom-cavity regime, entangled states
with two photons can be created from the vacuum state $|g,0\rangle $ under
the corresponding resonance, analogously to the dynamical Casimir effect in
a vibrating cavity containing a resonant two-level atom.

We also showed that the photon generation from vacuum occurs for a small
nonperiodic time modulation of the atom-cavity coupling, the atomic
transition frequency, or both in circuit QED. We deduced an effective
Hamiltonian for the bichromatic modulation in the dispersive regime, and
demonstrated that it describes well the exact dynamics. Numerical
simulations confirmed that several photons can be generated from vacuum
provided the modulation frequencies are fine tuned adiabatically.

This study illustrates the importance of the anti-rotating terms in the Rabi
Hamiltonian, ignored in the Jaynes-Cummings model -- here this term is
responsible for photon generation from vacuum and field amplification.
Moreover, one could engineer effective interactions in nonstationary circuit
QED by means of modulating the system parameters. As applications, this
scheme can be used to verify photon creation from vacuum in nonstationary
circuit QED due to an analog of the dynamical Casimir effect using a single
atom, as well as off-resonant population transfer between the states $%
\left\{ |g,n\rangle ,|e,n\pm 1\rangle \right\} $ and generation of entangled
states with several photons.

\begin{acknowledgments}
Work supported by FAPESP (SP, Brazil, contract No. 04/13705-3), ITAMP (at
Harvard University and Smithsonian Astrophysical Observatory) and
(partially) by National Science Foundation.
\end{acknowledgments}

\end{document}